\documentclass[10pt, aps,prc,twocolumn,superscriptaddress,preprintnumbers,
amsmath, 
floatfix,
longbibliography,
nofootinbib
]{revtex4-1}
\usepackage[T1]{fontenc}
\usepackage[utf8x]{inputenc} 
\usepackage{adjustbox}          
\usepackage[caption=false]{subfig}
\usepackage{url}
\usepackage{color}
\usepackage{float}
\usepackage[pdftex,colorlinks=true, linkcolor = blue, citecolor=blue,urlcolor=blue, bookmarksnumbered=true, bookmarksopen=true]{hyperref}
\usepackage{longtable}
\usepackage{amsfonts}
\usepackage{wrapfig,bm}
\newcommand{\beq}{\begin{equation}}
\newcommand{\eeq}{\end{equation}}
\newcommand{\bea}{\begin{eqnarray}}
\newcommand{\eea}{\end{eqnarray}}

\begin{document}

\title{Projection of Good Quantum Numbers for Reaction Fragments}
  
\author{Aurel Bulgac}%
\email{bulgac@uw.edu}%
\affiliation{Department of Physics,%
  University of Washington, Seattle, Washington 98195--1560, USA}
  
\date{\today}


\begin{abstract}

In reactions the wave packets of the emerging products typically are not eigenstates of particle number 
operators or any other conserved quantities and their properties are entangled.
I describe a particle projection technique in parts of space, which eschews the need to evaluate 
Pfaffians in the case of overlap of generalized Slater determinants or Hartree-Fock-Bogoliubov type of  vacua.
The extension of these formulas for calculating either angular momentum or particle projected energy 
distributions of the reaction fragments are presented as well.  The generalization to simultaneous particle and 
angular momentum projection of various reaction fragment observables is straightforward.
\end{abstract}

\preprint{NT@UW-19-8}

\maketitle

\section{\bf Introduction}

In practice sometimes one is interested in decomposing a many-nucleon 
wave function restricted to a part of the space 
into components with integer number of fermions, 
as typically the fragment wave function is not characterized by 
a good particle number. 

Even if two initial colliding nuclei are characterized by good particle and other good quantum numbers, 
the emerging reaction fragments are a superposition of nuclei with many possible quantum numbers 
allowed by conservation laws.  When the fragments are so far apart after the collision that any interaction 
between them is negligible, by performing a measurement of the particle composition of one fragment
at once leads to a well defined particle number in the other fragment, in an obvious generalization 
of \textcite{Einstein:1935}  ``spooky action at a distance.'' But unlike in the case of Schr\"{o}dinger's cat,
in this case there are more than two possible outcomes. The situation becomes even more complex when there are 
more than two fragments in the final state.
Reaction fragments after exchanging particles, energy, angular momenta, $\dots$ emerge entangled.

A simple example is that of the collision of a hydrogen atom 
with a positively charged naked ion.  After the collision, when the proton and the ion are infinitely separated,  
one can find the electron 
wave function fragmented between the potential well of the initial hydrogen atom and the potential well of the initially naked ion. 
If one were to make a measurement of where the electron is, one would find it present with different probabilities 
either attached to the proton, to the ion, or even to a free electron. These probabilities are straightforward to evaluate as the integral 
$\int_{P,I} \!d{\bf r} |\phi({\bf r})|^2$ over either the proton ($P$) or ion ($I$) region of space will give the 
probabilities to find the electron attached to either the proton or the ion. If these probabilities do not add up to one 
that would tell us that the electron has been ejected with some finite probability.

In the case of a many-particle system the evaluation of the probability to find an integer particle number in either
of the emerging  nuclear systems is a bit more convoluted, and that will be discussed in this paper, with  the emphasis on the 
case of the collision of partners with pairing correlations.
The relatively simple example of the collision of ``two hydrogen 
atoms'' is discussed in the next section.
A more complicated case is that of two nuclei colliding within the Hartree-Fock approximation and that was considered in 
Ref.~\cite{Simenel:2010} and it will be reviewed in the next section.
After the collision the receding wave packets are typically not characterized by good particle numbers. 
The initial target and projectile nuclei can be described by non-overlapping Slater determinants. However, 
after they come into contact the single particle wave functions of the initially separated nuclei evolve in the common
mean field of the combined nuclear system. Upon separation the projectile and target like nuclei end up with 
different number of nucleons and the 
single-particle wave functions of the initially separated partners are fragmented 
with components present in both emerging nuclear systems, and some nucleons might be even knocked out.

In section \ref{sec:II} I will review the particle projection in the case of normal nuclei (no pairing correlations), which is the
limiting case of the superfluid nuclei when pairing gap vanishes. In section  \ref{sec:III} I will  present the case of superfluid nuclei treated 
in the Bardeen-Cooper-Schrieffer
(BCS) approximation, which is formally equivalent to treating pairing correlation in the canonical 
basis~\cite{Bloch:1962,Balian:1969, Ring:2004}. The projection in the case of generalized 
Slater determinants is described in section \ref{sec:IV}.
Extensions of the projection method introduced in this paper to angular momentum distributions 
and particle projected energy are described in a somewhat 
brief, although full manner, in sections \ref{sec:V} and \ref{sec:VI}. As this is a relatively short paper, the main conclusions 
from the abstract, and others sections  are not reiterated again at the end.

\section{\bf Projecting the  particle number in part of the space in the case of a single Slater determinant} \label{sec:II}

I assume that the space has two (or more) partitions, 
the left ($z<0$) and the right ($z>0$) half-spaces, characterized by the corresponding  particle number operators
$\hat{N}_{L,R}$
\begin{align}
&\hat{N}_{L,R}=\int d\xi \Theta(\mp z)\psi^\dagger(\xi)\psi(\xi) 
 \end{align}
where $\psi^\dagger(\xi)$ and $\psi(\xi)=\sum_na_n\phi_n(\xi)$ are field operators, 
$\phi_n(\xi ) =\langle \xi|n\rangle = \langle0| \psi(\xi) a^\dagger_n|0\rangle $, $|0\rangle$ is the vacuum state, 
$\xi={\bf r},\sigma$ stands for spatial ${\bf r}=(x,y,z)$, spin $\sigma=\uparrow,\downarrow$, and isospin $\tau=n, p$ coordinates, 
$\Theta(z)$ is the Heaviside function, and the integral stands for the integral  
over spatial coordinates and the summation of spin coordinates. 
$a^\dagger_n$ and $a_n$ are the creation and annihilation operators  for single particle states with wave functions $\phi_n(\xi)$.
The total average numbers of particles in 
the left and right half-spaces are naturally given by the expressions
\begin{align}
N_{L,R}= \int d\xi  \Theta(\mp z) \sum_{n=1}^A  |\phi_n(\xi)|^2 ,
\end{align}
where the sum is over occupied single-particle states.
In the subsequent formulas one should make a distinction between 
the operator $\hat{N}$ and its respective expectation values $N$.
Obviously, one can separate the entire space in arbitrary ways, 
e.g. the interior and the exterior of a sphere.

The particle projectors on half-space $L$ and $R$ are
\begin{align}
&\hat{P}_{L,R}(N)= \int_{-\pi}^\pi \frac{d\eta}{2\pi} e^{i\eta(\hat{N}_{L,R}-N) }, \label{eq:proj}\\
&e^{i\eta \hat{N}_{L,R} }=1+\Theta(\mp z) (e^{i\eta \hat{N}_{L,R}}-1), \label{eq:exp}\\
&\langle \phi_n | e^{i\eta\hat{N}_{L,R} } | \phi_m \rangle =\delta_{nm}+(e^{i\eta}-1)\langle \phi_n|\Theta(z)|\phi_m\rangle.
\end{align}
Equation \eqref{eq:exp} is obtained by expanding the exponential and using $\Theta^2(\mp z)\equiv \Theta(\mp z)$.
The probability $P_R(N)$  to find exactly $N$ particles in the right half-space is given by~\cite{Simenel:2010}
\begin{align}
&P_R(N)=\langle \Phi |\hat{P}_R(N)|\Phi\rangle = \int_{-\pi}^\pi \frac{d\eta}{2\pi} e^{-i\eta N}  \langle \Phi|\Phi(\eta)\rangle\\
&\text{where} \quad |\Phi\rangle = \prod_{n=1}^A a^\dagger_n |0\rangle,\\
&\langle \Phi | e^{i\eta\hat{N}_R}|\Phi\rangle = \langle \Phi|\Phi(\eta)\rangle =\det \left (\delta_{mn} +O_{mn}(\eta)\right ) ,\\ 
&O_{nm}(\eta)= (e^{i\eta}-1)\langle \phi_n |\Theta(z) |\phi_m\rangle, \label{eq:det0}.
\end{align}
The action of the operator $e^{i\eta \hat{N}_{L,R}}$ on a Slater determinant
is equivalent to a fictitious time-dependent evolution of the single-particle states 
in an external field only $\Theta(\mp z)$ and therefore 
\begin{align}  
&\phi_n(\xi,\eta)= [1+\Theta(\mp z)(e^{i\eta}-1)]\phi_n(\xi),
\end{align}
where I used the relation $e^{i\eta\Theta(\mp z)}= 1 + \Theta(\mp z)(e^{i\eta}-1)$.

By diagonalizing at first the matrix $\langle \phi_n|\Theta(z)|\phi_m\rangle$ the numerical calculations are greatly simplified.
If the eigenvalues of the overlap matrix $\langle \phi_n|\Theta(z)|\phi_m\rangle$, which is Hermitian positive semi-definite,  
are $0\le \alpha_n\le 1$, then
\begin{align}
P_{R}(N)= \int_{-\pi}^\pi \frac{d\eta}{2\pi} e^{-i\eta N}\prod_{n=1}^A\left [1+ (e^{i\eta}-1)\alpha_n\right ], \label{eq:P_N}
\end{align} 
and similar formulas for the particle number probability $P_L(N)$. Note also that $N_R=\sum_n \alpha_n$. 
Obviously, the following relations hold
\begin{align}
&P_L(N)=P_R(A-N),\\
&\sum_{N=0}^AP_{L,R}(N)=1.
\end{align}
Note that this formula does not explicitly reveal if nucleons have been knocked out and are not attached to either fragment. 
It is however straightforward to generalize the present formulas to account for emitted nucleons or even clusters.   

To illustrate the formalism, let me consider here an idealized case of the collision of two ``hydrogen atoms,'' each initially with an electron 
in its respective ground state when they are infinitely separated ($z \rightarrow \infty$ for $t\rightarrow -\infty$). 
The ``nuclei'' will follow a classical trajectory and only the 
``electrons are treated quantum mechanically. The initial ``electronic'' wave function is a Slater determinant of two orthonormal 
single-particle wave functions
\begin{align}
&\phi_\pm({\bf r},t)= \frac{1}{2} \left [ \phi({\bf r}-{\bf z},t) \pm \phi({\bf r}+{\bf z},t)\right ],\\
&\int d^3r |\phi({\bf r},t)|^2 = 1,  \\&
\lim_{t\rightarrow {-\infty} } \int d^3r \phi({\bf r}-{\bf z},t)  \phi^*({\bf r}+{\bf z},t) = 0,
\end{align}
where ${\bf z}=(b,0,z)$ and $2b$ is the impact parameter.
 After the collision the Slater determinant will have a similar structure and the overlap matrix Eq. \eqref{eq:det0} will be
 \begin{align}
 {\cal O}(\eta) = \frac{e^{i\eta}-1}{2}\left(  \begin{array}{rr} 1& -1\\
 -1&1\\\end{array} \right ),
 \end{align}
 which,  after using Eq. \eqref{eq:P_N}, will lead to exactly one particle per ``nucleus'' in the final state, 
 as one would naturally expect in this case. \\
 
\section{\bf Projecting the particle number in the case of  generalized Slater determinants in the canonical basis.}\label{sec:III}

In the case when pairing correlations are present the nucleus wave function in the canonical basis is given by
\begin{align}
| \Phi \rangle = \prod_{n=1}^\Omega(u_n+v_n a^\dagger_n a^\dagger_{\bar{n}}) | 0 \rangle , \label{eq:Phi}
\end{align}
where $u_n^2+v_n^2=1$ and $n$ and $\bar{n}$ denote time-reverse single-particle states. 
In order to project the particle number one introduces 
a rotated in the gauge space wave function, in which case  
\begin{align}
& u_n\rightarrow u_n, \quad v_n\rightarrow e^{2i\eta}v_n, \label{eq:vn} \\
&| \Phi (\eta)\rangle = e^{i\eta\hat{N}}|\Phi\rangle=\prod_{n=1}^\Omega (u_n+e^{2i\eta}v_n a^\dagger_n a^\dagger_{\bar{n}}) | 0 \rangle ,\label{eq:BCS0}\\
& \Phi_N\propto \int_{ -\tfrac{\pi}{2} }^{ \tfrac{\pi}{2} } \frac{d\eta}{\pi}e^{-i\eta N} |\Phi(\eta)\rangle  \propto 
\left ( \sum_n  \frac{v_n}{u_n} a^\dagger_n a^\dagger_{\bar{n}} \right )^{\tfrac{N}{2}},
\end{align}
where 
\begin{align}
\hat{N}=\sum_{n=1}^\Omega (a_n^\dagger a_n+a_{\bar n}^\dagger a_{\bar n}),
\end{align}
and $\Phi_N$  has exactly $N$-particles and
\begin{align}
|\Phi(\eta)\rangle = \sum_{N=0}^{2\Omega}e^{i\eta N} a_N|\Phi_N\rangle
\end{align}
where $|N\rangle$ have fixed particle number and only even $N$ particle states contribute to the sum.

In an unitary transformation generated by the operator $e^{i\eta\hat{N}}$ 
one would consider instead $u_n\rightarrow e^{-i\eta}u_n, \quad v_n\rightarrow e^{i\eta}v_n,$ 
which would lead to a total wave function with a different overall phase
$|\tilde{ \Phi} (\eta)\rangle = e^{-i\eta\Omega}\prod_{n=1}^\Omega (u_n+e^{2i\eta}v_n a^\dagger_n a^\dagger_{\bar{n}}) | 0 \rangle. $
For a zero-range interaction one should take the limit $\Omega\rightarrow \infty$, or at least consider $\Omega\gg A$, in which case 
$\sum_{n=1}^\Omega u_nv_{\bar{n}}^*\rightarrow \infty$ 
for $\Omega\rightarrow \infty$~\cite{Bulgac:2002} and use the appropriate regularization and renormalization procedures for calculations.

The overlap $\langle\Phi|\Phi(\eta)\rangle=\prod_n\left ( |u_n|^2 +e^{2i\eta}|v_n|^2\right )$ is a periodic function with period $\pi$ and hence
the probability  to find exactly $N$ particles (as when $N$ is even and there are $N/2$ pairs)  is given by the Fourier transform of 
$\langle \Phi| | \Phi(\eta) \rangle $
\begin{align}
P(N)=\int_{ -\tfrac{\pi}{2} }^{ \tfrac{\pi}{2} } \frac{d\eta}{\pi}e^{-i\eta N}\prod_{n=1}^\Omega \left [ 1 + (e^{2i\eta}-1)  |v_n|^2\right ]. \label{eq:BCS1}
\end{align}
Notice that the integrand vanishes iff $\eta =\pm \tfrac{\pi}{2}$ and at least 
for one $n$ also  $|v_n|^2\equiv \tfrac{1}{2}$, thus never inside the integration interval.

\begin{figure}
\includegraphics[width=0.9\columnwidth]{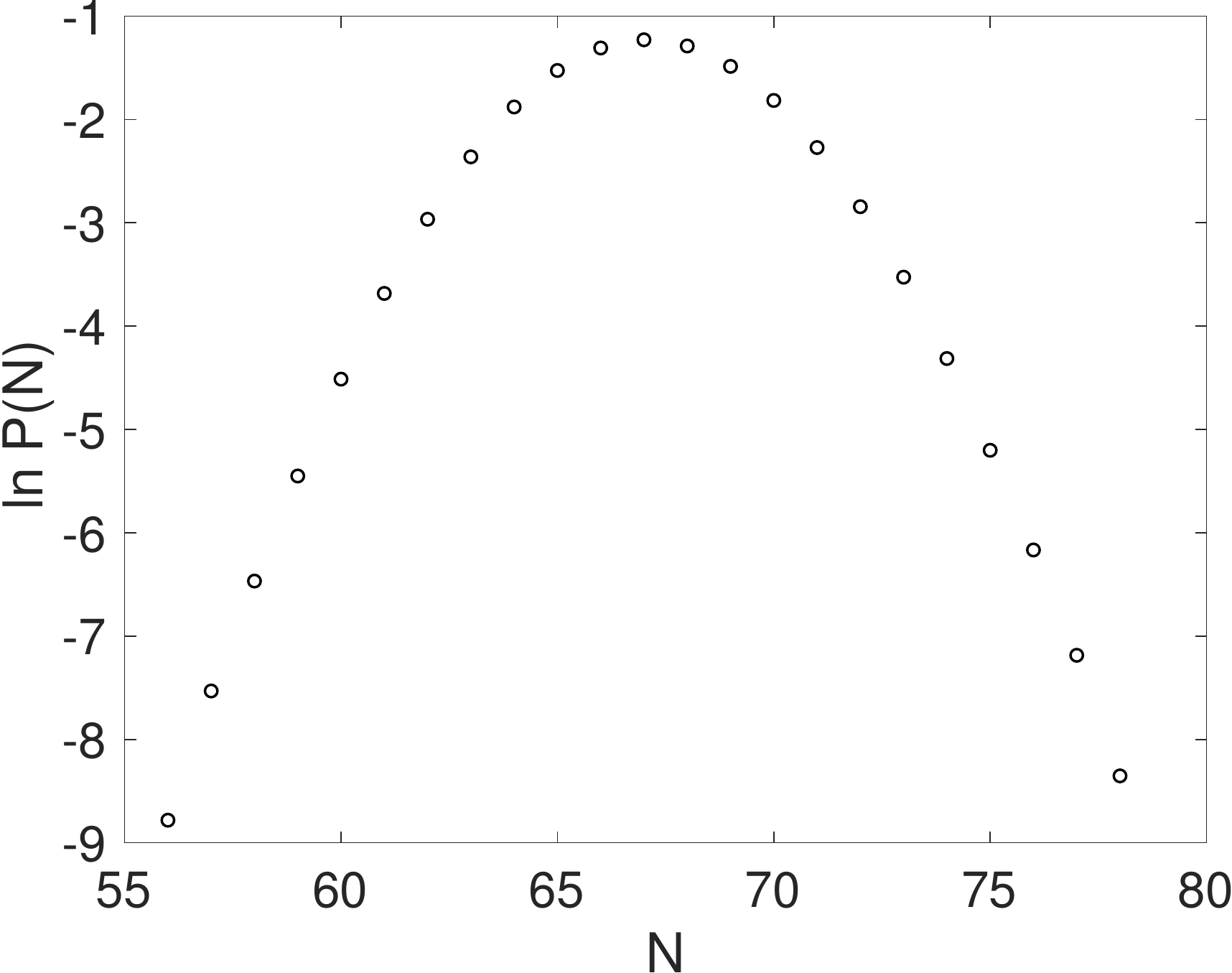}
\caption{\label{fig:Z1}The probability $P(N)$ of finding $N$ particles in a BCS wavefunction, derived from the function  $Z(\eta)$ in Fig. \ref{fig:Z}.}
\end{figure}

\begin{figure}
\includegraphics[width=0.9\columnwidth]{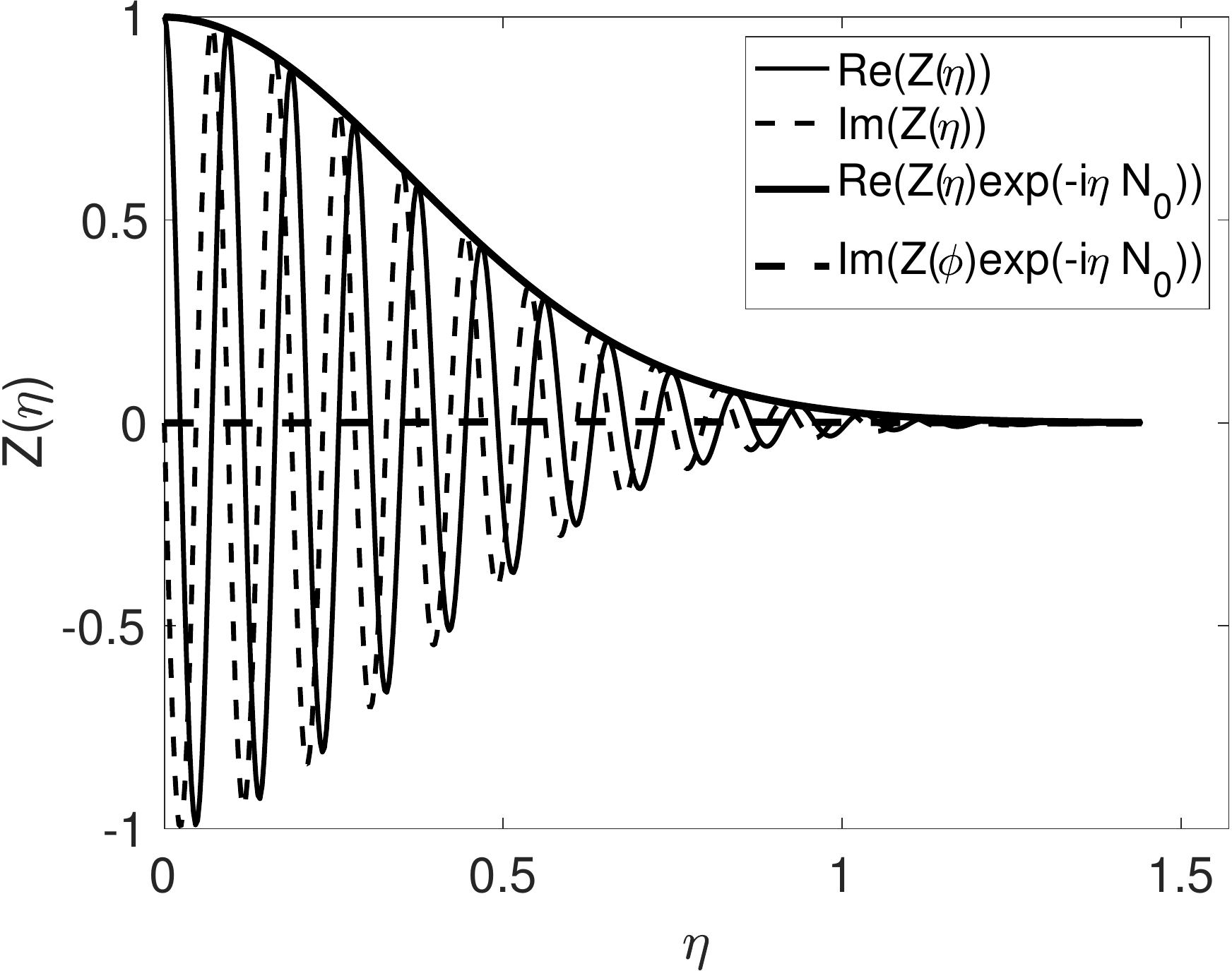}
\includegraphics[width=0.9\columnwidth]{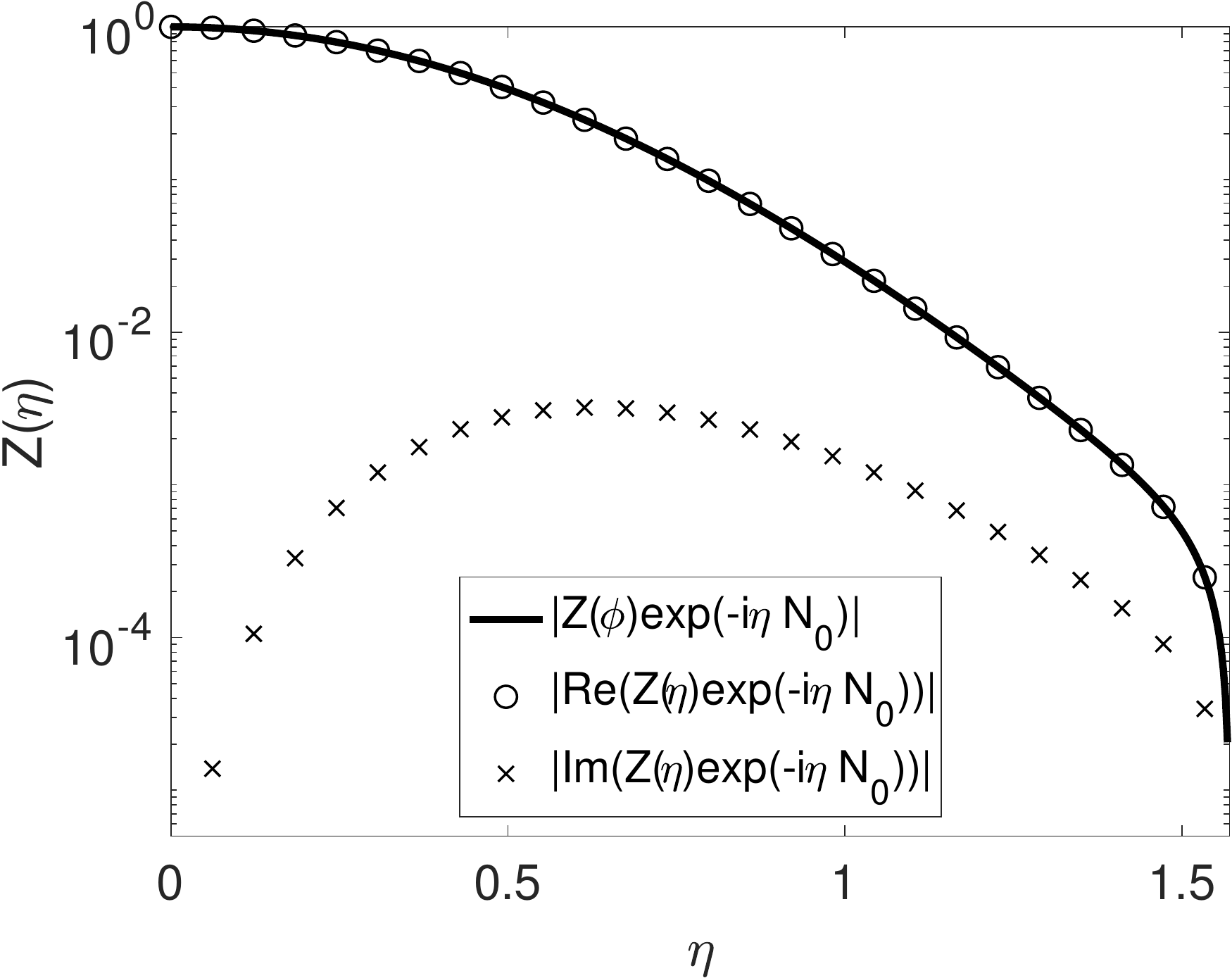}
\caption{\label{fig:Z}The function $Z(\eta)$, with the factor $\exp(-i\eta N)$ included or not, evaluated for 
$|v_n|^2=\tfrac{1}{2}\left [1-\tfrac{\varepsilon_n-\mu}{\sqrt{ (\varepsilon_n-\mu)^2+\Delta^2} }\right ]$, 
for a uniform single-particle spectrum $\varepsilon_n\propto n$,  $n=1,2,...,\infty$, $\mu=34$, and $\Delta=2.5$. }
\end{figure}

One can show, by explicit numerical evaluation, that the imaginary part of the quantity $\tilde{Z}(\eta)$
\begin{align}
&Z(\eta) = \prod_{n=1}^\Omega\left [1+ (e^{2i\eta}-1)|v_n|^2\right ], \\
&\tilde{Z}(\eta) = e^{-i\eta N_0} Z(\eta),  \label{eq:Z}
\end{align}
with $N_0=\langle \Phi |\hat{N}|\Phi\rangle = 2\sum_{n=1}^\Omega |v_n|^2$,
is orders of magnitude smaller than its real part, if $|v_n|^2$ has a Fermi-like thermal or BCS-like shape 
and $|v_n|^2\rightarrow 0$ when the upper limit $\Omega\rightarrow \infty$, see Fig. \ref{fig:Z1}. 
$\tilde{Z}(\eta)$ is basically a non-oscillatory function and 
$|\tilde{Z}(\eta)|\approx \tilde{Z}(\eta)$ has a bell shape around $\eta=0$ and may vanish only at $\eta = \pm \tfrac{\pi}{2}$.

The numerical evaluation of Eq. \eqref{eq:BCS1} then becomes much simpler, see Fig. \ref{fig:Z},  and much 
more accurate over orders of magnitude, as one can instead evaluate
 \begin{align}
P(N)= \int_{-\tfrac{\pi}{2} }^{ \tfrac{\pi}{2} }\frac{d\eta}{\pi} e^{-i\eta(N-N_0)}\tilde{Z}(\eta) \label{eq:PR}
\end{align} 
for a quite large number of values of $N$ around the mean value 
$N_0$, using a relatively small number of quadrature points, 
after establishing that the integrand is not a fast oscillating function of $\eta$ for $N$ very different from $N_0$. 
The additional factors $e^{-iN\eta}$ and $e^{iN\eta}$ 
cancel in Eq. \eqref{eq:PR} and were introduced only to reveal the properties of the integrand.   
Since the integral is real the formula can be simplified
\begin{align}
P_{R}(N)= 2\; {\text Re}\int_{0}^{\tfrac{\pi}{2}} \frac{d\eta}{\pi} e^{-i\eta N}Z(\eta). 
\end{align}

\vspace{0.5cm}

\section{\bf Particle projection in part of the space in the case of a generalized Slater determinant} \label{sec:IV}

During time evolution initially time-reversed single-particle states
in general cease to satisfy time-reversal symmetry, e.g. in the presence of a time-dependent external magnetic field,  
and in that case one should 
use the more general formulas below, see Eq. \eqref{eq:PSN1}.
In Eqs. \eqref{eq:Phi}, \eqref{eq:BCS0}, and \eqref{eq:BCS1} above no projection on a ``half''-space is implied. 

When discussing time-dependent problems, 
in particular well separated spatially fission fragments, the most convenient representation is in the real space, 
which I explicitly recapitulate here.
The creation and annihilation quasi-particle operators are represented as~\cite{Ring:2004}
\begin{align}
\!\!\!\!\!\! &\alpha_k^\dagger  = 
\int d\xi\left [ {\textrm u}_k(\xi) \psi^\dagger (\xi) + {\textrm v}_k(\xi) \psi (\xi)\right ], \label{eq:a0}\\
\!\!\!\!\!\! &\alpha_k= 
\int d\xi\left [ {\textrm v}_k^*(\xi) \psi^\dagger (\xi ) + {\textrm u}_k^*(\xi) \psi (\xi)\right ], \label{eq:b0}
\end{align}
and the reverse relations
\begin{align} 
&\psi^\dagger (\xi) = \sum_k \left [ {\textrm u}^*_k(\xi)  \alpha^\dagger _k  
                                            + {\textrm v}_k(\xi)\alpha_k \right ], \label{eq:p1}\\
&\psi(\xi) =                \sum_k \left [ {\textrm v}^*_k(\xi)\alpha^\dagger_k
                                            + {\textrm u}_k(\xi)\alpha_k \right ], \label{eq:p2}
\end{align}
where $\psi^\dagger (\xi )$ and $ \psi (\xi)$ are the field operators 
for the creation and annihilation of a particle with coordinate $\xi$.
The normal number (Hermitian $n=n^\dagger$ ) and anomalous (skew symmetric $\kappa =-\kappa^T$) densities are
\begin{align}
&n(\xi,\xi') = \langle 0|\psi^\dagger(\xi')\psi(\xi)|0\rangle = \label{eq:number} \\
& \sum_k {\textrm v}_k^*(\xi) {\textrm v}_k(\xi')=\sum_{l=n,\bar{n}} v_l^2  \phi_l^*(\xi) \phi_l(\xi'),\nonumber \\
&\kappa(\xi,\xi') = \langle 0|\psi(\xi')\psi(\xi)|0\rangle = \\
&\sum_k {\textrm v}_k^*(\xi){\textrm u}_k(\xi') = \sum_{l=n,\bar{n}} u_lv_l \phi_l^*(\xi) \phi_{\bar{l}}^*(\xi'), \nonumber \\
&  \int d\xi \phi_k^*(\xi) \phi_l(\xi)=\delta_{kl},
\end{align}
with $u_l^2+v_l^2=1$, $0\le u_l=u_{\bar{l}}\le 1$,  $0\le v_l=-v_{\bar{l}}\le 1$, and $n$ and $\bar{n}$ denote time-reversed states 
in the canonical representation~\cite{Bloch:1962,Ring:2004,Balian:1969}, and where
\begin{align}
& \alpha_k|\Phi\rangle =0, \quad \langle \Phi|\alpha_k^\dagger = 0, \quad \langle \Phi|\alpha_k\alpha_l^\dagger|\Phi\rangle =\delta_{kl}.
\end{align}

In the case of a generalized Slater determinant $|\Phi \rangle$ 
the  total wave function rotated in the gauge space is obtained
in a similar manner to the Hartree-Fock case discussed above 
on projecting on particle number on the right half-space. 
At this point I will introduce new kinds of creation and annihilation quasiparticle operators.
The result of the 
``gauge'' rotation  on the quasiparticle wave functions, which leads to a similar transformation to Eq. \eqref{eq:vn}, is defined as
\begin{align}
&u_n(\xi,\eta) = u_n(\xi),\quad  v_n(\xi,\eta) = e^{2i\eta\Theta(z)}v_n(\xi), \label{eq:uv}
\end{align}
which leads to the new type of creation and annihilation operators 
\begin{align}
&\alpha_k^\dagger(\eta) = 
\int d\xi\left [ {\textrm u}_k(\xi) \psi^\dagger (\xi) + e^{2i\eta\Theta(z)}{\textrm v}_k(\xi) \psi (\xi)\right ],\label{eq:a}\\
&\alpha_k(\eta)= 
\int d\xi\left [ e^{-2i\eta\Theta(z)}{\textrm v}^*_k(\xi) \psi^\dagger (\xi) +  {\textrm u}^*_k(\xi) \psi (\xi)  \right ].\label{eq:b}
\end{align}
The anti-commutation relations for these operators are
\begin{align}
\!\!\!\!\! &\{\alpha^\dagger_k(\eta),\alpha^\dagger_l(\eta)\}=B_{kl} \\
&=\int d\xi  
\left [ {\textrm u}_k(\xi) {\textrm v}_l(\xi)  + {\textrm v}_k(\xi) {\textrm u}_l(\xi)\right ]e^{2i\eta\Theta(z)},\nonumber  \\
\!\!\!\!\! &\{\alpha^\dagger_k(\eta),\alpha_l(\eta)\}=\delta_{kl}. \label{eq:BB}
\end{align}
These operators are similar to the operators obtained with non-unitary transformations by 
\textcite{Balian:1969}, which preserve Eq. \eqref{eq:BB}.
By inserting Eqs. (\ref{eq:p1}) and  (\ref{eq:p2}) into Eqs. (\ref{eq:a}) and (\ref{eq:b})  one can establish that
\begin{align}
&\alpha_k^\dagger(\eta) = \sum_l A_{kl}    (\eta)\alpha^\dagger_l + B_{kl}   (\eta)\alpha_l, \label{eq:a1} \\
&\alpha_k(\eta)               = \sum_l B_{kl}^*(\eta)\alpha^\dagger_l + A_{kl}^*(\eta)\alpha_l, \label{eq:b1}
\end{align}
where these matrices are 
\begin{align}
& A_{kl}(\eta)= \int d\xi \left [ {\textrm u}_k(\xi) {\textrm u}^*_l(\xi) + {\textrm v}_k(\xi){\textrm v}_l^*(\xi) e^{ 2i\eta\Theta(z)} \right ] \label{eq:a2} \\
&  =\delta_{kl} +  (e^{2i\eta}-1) \int d\xi  \Theta(z){\textrm v}_k(\xi){\textrm v}_l^*(\xi),  \label{eq:a20}\\
& B_{kl}(\eta)= \int d\xi \left [ {\textrm u}_k(\xi) {\textrm v}   _l(\xi) + {\textrm v}_k(\xi){\textrm u}_l   (\xi) e^{2i\eta\Theta(z)} \right ]\label{eq:b2}\\
& = (e^{2i\eta}-1)\int d\xi  \Theta(z){\textrm v}_k(\xi){\textrm u}_l (\xi) = B_{lk}(\eta). \label{eq:b20}
\end{align}
In deriving Eqs. (\ref{eq:a20}) and (\ref{eq:b20}) I took into account that the transformation from field to quasiparticle operators is unitary.

Using the technology described 
by \textcite{Balian:1969}, \textcite{Ring:2004} in Appendix E, and particularly 
the method introduced by \textcite{Mizusaki:2018}  one can show that
\begin{align}
&|\Phi(\eta)\rangle = {\cal N} (\eta) e^{\hat{Z}(\eta)} |\Phi\rangle, \quad \langle \Phi(\eta)|\Phi(\eta)\rangle =1, \label{eq:wf}\\
&\text{where}\quad \hat{Z}(\eta)=  \sum_{k<l} [A(\eta)^{-1}B(\eta)]^{*}_{kl}\alpha_k^\dagger\alpha_l^\dagger, \label{eq:Z1}\\
& \alpha_k(\eta) |\Phi(\eta)\rangle = 0, \quad  \langle \Phi | \Phi (\eta) \rangle  = \sqrt{\det A}= {\cal N} (\eta),
\end{align}
with ${\cal N}(0)=1$ and  the last relation is  known as
the Onishi and Yoshida formula~\cite{Onishi:1966, Balian:1969, Ring:2004,Mizusaki:2018}. 
Note that only the antisymmetric part of the matrix $A(\eta)^{-1}B(\eta)$ is contributing in Eq. \eqref{eq:Z1} to the operator $\hat{Z}(\eta)$.

The overlap $\langle \Phi|\Phi(\eta)\rangle $ becomes in this case
\begin{align}
& \langle \Phi | \Phi (\eta) \rangle =  \sqrt{ \det{\left [ \delta_{kl} +(e^{2i\eta}-1)  {\cal O}_{kl} \right ] }} , \label{eq:det}\\ 
&{\cal O}_{kl}= \langle {\textrm v}_l|\Theta(z)|{\textrm v}_k\rangle, \label{eq:O}
\end{align}
and where ${\textrm v}_{k,l}$'s are the ${\textrm v}$-components of the quasiparticle wave functions and 
the indices $k$ and $l$ run over all single-particle states, e.g. both $n$ and its counterpart $\bar{n}$, 
for which ${\textrm u}_n={\textrm u}_{\bar{n}}$ and ${\textrm v}_n=-{\textrm v}_{\bar{n}}$ in the representation where
the number density $n(\xi,\xi')$ is diagonal and the anomalous density $\kappa(\xi,\xi')$ 
is anti-symmetric $2\times2$ block-diagonal~\cite{Bloch:1962,Ring:2004}.
For stationary states there is no sign ambiguity in choosing the sign of the square root in Eq. \eqref{eq:det}~\cite{Sheikh:2000}.
Again, since the matrix 
${\cal O}$ is Hermitian (and positive semi-definite) it can be diagonalized. 

The probability to find $N$ particles in the right half-space is given in this case by
\begin{align}
P_R(N)=\int_{-\tfrac{\pi}{2}}^{\tfrac{\pi}{2}} \frac{d\eta}{\pi} e^{-i\eta N}
\sqrt{ \prod_{l=1}^{2\Omega}  \left [1+(e^{2i\eta}-1)\beta_l\right ]  }. \label{eq:PSN}
\end{align}
where $0\le\beta_l\le 1$ are the eigenvalues of overlap matrix ${\cal O}$, see Eq. \eqref{eq:O}, and 
\begin{align}
&\langle \Phi | \Phi (\eta) \rangle  
=\sqrt{ \prod_{l=1}^{2\Omega}  \left [1+(e^{2i\eta}-1)\beta_l\right ] }.
\end{align}
A factor $\left [1+(e^{2i\eta}-1)\beta_l\right ]$ is zero if and only if  
both $\eta\equiv \pm\tfrac{\pi}{2}$ and $\beta_l\equiv \tfrac{1}{2}$, thus exactly at the upper and lower limits of the integration interval only. 
Therefore there is no ambiguity in this case as well for choosing the sign of the square root. 

The above formulas can be simplified a little bit further, as  $P_R(N)$ is 
real and Eq. \eqref{eq:PSN}
can be reduced to
\begin{align}
\!\!\!\!\!\!\!P_R(N)=2\;{\text Re}\int_0^{\tfrac{\pi}{2}} \frac{d\eta}{\pi} e^{-i\eta N}\sqrt{ \prod_{l=1}^{2\Omega} \left [1+(e^{2i\eta}-1)\beta_l\right ] }.
\label{eq:PNR}
\end{align}
With the replacement $\Theta(z)\rightarrow 1$
and after the diagonalization of the matrix $\langle {\textrm v}_l|{\textrm v}_k\rangle$
one recovers the canonical basis result, see Eq. \eqref{eq:BCS1}.
The particular case when the generalized Slater determinant $|\Phi\rangle$ is 
represented in the canonical basis, and when the states $n$ and $\bar{n}$ do not anymore satisfy the time-reversal symmetry, 
follows from Eq. \eqref{eq:PSN}. The comments made above, see Eqs. \eqref{eq:Z} and \eqref{eq:PR}, about the 
oscillatory character of the integrand apply here as well. In particular one also has $N_R = \sum_{l=1}^{2\Omega} \beta_l$.  And finally,
for the left half-space one obviously has $P_L(N) = P_R(A-N)$.

Note the difference among Eqs. \eqref{eq:det0} and \eqref{eq:BCS1}  (where there is no square 
root) and Eqs. \eqref{eq:det}, \eqref{eq:PSN}, and \eqref{eq:PNR} (where there is a square root). 
When pairing correlations vanish one naively expects that Eqs. \eqref{eq:det0} and \eqref{eq:BCS1} and 
\eqref{eq:det} and \eqref{eq:PSN}  should agree.
However, in the case of ordinary Slater determinants the projected value of $N$ and the dimension of the matrix ${\cal O}$ 
can be even or odd, and for that reason the integration interval on $\eta$ is $[-\pi,\pi]$.
For generalized Slater determinants the dimension of the matrix ${\cal O}$  is always even
and the integration interval is now $[-\pi/2,\pi/2]$. 
When there are degenerate time-reversal orbitals $n$ and $\bar{n}$, after extracting the square root in Eq. \eqref{eq:PSN} 
one is left with half the number of factors in the product, as in the case of Eq. \eqref{eq:BCS1}, where there 
is no square root. In Eq. \eqref{eq:BCS1} 
the product runs only over $n$ states, but not over their time-reversed partners $\bar{n}$.
Therefore Eq. \eqref{eq:BCS1} agrees with Eq. \eqref{eq:PSN} in the case when there are degenerate time-reversed orbitals.
One can project in this case only on even particle numbers $N$ in the right half-space. 
The generalization to a system with pairing correlations and total odd 
particle numbers is straightforward~\cite{Ring:2004}.

Recently \textcite{Mizusaki:2018} 
clarified the reasons why the Onishi and Yoshida formula does not have a sign ambiguity, 
particularly in the case when the size of the Fock space is finite, as is the case in 
the overwhelming majority of numerical implementations. They have proven that Onishi and Yoshida~\cite{Onishi:1966}
and Robledo~\cite{Robledo:2009} formulas for the norm overlaps are identical in this case.  
A different approach to evaluate number of particles in fission 
fragments was recently suggested by \textcite{Verriere:2019}.

There is a generalization of 
Eq. \eqref{eq:PSN} to the case when in the right half-space there are fragments with an odd particle number, which can happen  for example
when during time-dependent evolution Cooper pairs break up and partners in initially  time reversed orbitals can end up in different half-spaces.
This is achieved by replacing $2\eta\rightarrow \eta$ in Eq. \eqref{eq:uv}, which leads to obvious changes 
in the ensuing equations for both even and odd $N$ values and the integration interval changes to $[-\pi,\pi]$.  Namely
\begin{align}
& \langle \Phi | \Phi (\eta) \rangle  = \sqrt{ \det{\left [ \delta_{kl} +(e^{i\eta}-1)  {\cal O}_{kl} \right ] }} ,\\ 
&P_R(N)={\text Re} \int_{0}^{\pi} \frac{d\eta}{\pi} e^{-iN\eta}\sqrt{ \prod_l \left [1+(e^{i\eta}-1)\beta_l\right ] },\label{eq:PSN1}
\end{align}
where as before $0\le \beta_l\le 1$ are the eigenvalues of the matrix ${\cal O}$, which was defined in Eq. \eqref{eq:O}, 
but now $N$ can be both an even or an odd integer.
One can convinced oneself that there is no sign ambiguity in extracting the square root in this case either, 
following the same kind of  argument I presented above.

\section{\bf Extension to projecting the angular momentum} \label{sec:V}

In the case of three-diemnsional rotations one can develop a similar projection technique.
For simplicity let me consider a one-parameter group transformation, 
e.g. rotation around a single axis $\hat{R}(\eta)=e^{i\hat{J}_x\eta}$ perpendicular 
to the symmetry axis of a nucleus~\cite{Bertsch:2019}, 
and the corresponding  transformation of the components of the quasiparticle wave functions
\begin{align}
&{\textrm u}_n(\xi,\eta) \equiv {\textrm u}_n(\xi),\quad
{\textrm v}_n(\xi,\eta) =\hat{R}(\eta){\textrm v}_n(\xi). 
\end{align}   
Typically one would rotate both ${\textrm u}$- and ${\textrm v}$-components of the quasiparticle wave functions. Since one typically is 
interested only in the matter densities it is not necessary to rotate the ${\textrm u}$-components 
as well, similar to Eqs. \eqref{eq:a} and \eqref{eq:b}.   In this case the overlap matrix element 
$\langle \Phi|\Phi(\eta)\rangle $ is given by
\begin{align}
\langle \Phi | \Phi (\eta) \rangle  = \sqrt{ \det{ \left [ \delta_{kl} + 
\langle {\textrm v}_l |\hat{R}(\eta)| {\textrm v}_k\rangle - \langle {\textrm v}_l | {\textrm v}_k\rangle \right ] } }. 
\label{eq:rot}
\end{align}
In the canonical basis  the matrix $\langle {\textrm v}_l|{\textrm v}_k\rangle$ is 
diagonal and since $\hat{R}(\eta)=e^{iJ_x\eta} $,  
one can then prove that both matrices $\langle {\textrm v}_l | {\textrm v}_k\rangle $ 
and $\langle {\textrm v}_l |\hat{R}(\eta)| {\textrm v}_k\rangle$ can
 be diagonalized simultaneously.
Let me denote the eigenvalues of the matrices $\langle {\textrm v}_l | {\textrm v}_k\rangle $ 
and $\langle{\textrm v}_l |\hat{R(\eta)}| {\textrm v}_k\rangle$ 
with $v_n^2$ and  $w_n^2e^{i\lambda_n}$, respectively.   The sign of the overlap 
matrix element $\langle \Phi|\Phi(\eta)\rangle $ is ill defined for some values of $\eta$ if and only if  for
at least for one $n$ one has $w_n^2+v_n^2\equiv 1$ and at the same time 
inside the integration interval also $\lambda\equiv-\pi$.  
For example, in the case of nuclei invariant with respect to reflection symmetry 
$z\rightarrow -z$
a rotation by $\eta=\pm\pi$ leads to an identical state and in this case $w_n^2=v_n^2$ (though not necessarily to
$w_n^2+v_n^2\equiv 1$). However, the integration interval  for $\eta$ is $[-\pi,\pi]$ and the overlap matrix element vanishes exactly 
at the limits of the integration integral over $\eta$ iff $w_n^2+v_n^2\equiv 1$, 
and no ambiguity over the sign of the overlap matrix element $\langle \Phi|\Phi(\eta)\rangle $ arises in this case. The complex valued 
overlap $\langle \Phi|\Phi(\eta)\rangle $ is expected to be a continuous function of $\eta$ and sign ambiguities can
arise only if this overlap vanishes strictly inside the interval  $(-\pi,\pi)$. In the 
case of reflection symmetry $z\rightarrow -z$ it is sufficient to consider rotations only in the interval $[-\pi/2,\pi/2]$. 
In addition, axial symmetry in the presence of reflection symmetry $z\rightarrow -z$ 
also implies that one can reduce the integration interval even further to $[0,\pi/2]$.  

In the case of  the axially symmetric reaction fragments the individual probabilities can be evaluated using~\cite{Bertsch:2019}
\begin{align}
& |\Phi \rangle = \sum_{J}  a_{J} |J0\rangle,\\
&|a_J|^2 = (2J+1)\int_0^{\pi} \!\!\!d\eta \;\sin(\eta)\langle  \Phi |\Phi(\eta)\rangle P_J(\cos\eta),
\end{align}
where $|J0\rangle$ is the wave function with total angular momentum $J$ and $J_z=0$, and $P_J$ is a Legendre polynomial.

\section{\bf Extension to projecting  the particle number for other observables}\label{sec:VI}

One can define generalized density matrices
\begin{align}
&n(\xi,\xi'|\eta) = \langle \Phi|\psi^\dagger(\xi')\psi(\xi)|\Phi(\eta)\rangle, \label{eq:n1}\\
&\kappa(\xi,\xi'|\eta) = \langle \Phi|\psi(\xi')\psi(\xi)|\Phi(\eta)\rangle,\label{eq:n2} \\
&n_2(\xi_1,\xi_2,\zeta_1,\zeta_2|\eta) = \label{eq:n3}\\
& \langle \Phi|\psi^\dagger(\xi_1)\psi^\dagger(\xi_2)\psi(\zeta_2)\psi(\zeta_1)|\Phi(\eta)\rangle,\nonumber 
\end{align}
with $\Phi(\eta)\rangle$ defined in Eq. \eqref{eq:wf}. 
These generalized density matrices are well defined and have no divergencies. 
Using the Cauchy-Schwarz inequality it follows immediately that
\begin{align}
& |n(\xi,\xi|\eta)|^2     \le  \langle  \Phi|\psi^\dagger(\xi)\psi(\xi)|\Phi\rangle
                                       \langle \Phi(\eta)|\psi^\dagger(\xi)\psi(\xi)|\Phi(\eta)\rangle \nonumber                                  
\end{align}
and similar relations for $\kappa(\xi,\xi'|\eta)$ and $n_2(\xi_1,\xi_2,\xi_1,\xi_2|\eta)$.
General rules for evaluating such densities have been derived 
many times, see e.g. Refs.~\cite{Balian:1969,Ring:2004,Robledo:2009,Hu:2014,Avez:2012,Mizusaki:2013,Bertsch:2012}. 
Alternatively, one can invert Eqs. (\ref{eq:a}, \ref{eq:b}) to express $\psi^\dagger(\xi)$ and $\psi(\xi)$
in terms of $\alpha_k^\dagger(\eta)$ and $\alpha_k(\eta)$ and subsequently 
use $\langle \Phi| \alpha_k^\dagger =0 $ and 
$\alpha_k(\eta)|\Phi(\eta\rangle =0$.

In the case of a 
density functional theory approach the total energy of a system is a function(al) of various 
densities ${\cal E}[n(\xi,\xi), \ldots]$, where the ellipses stand for other densities not explicitly shown.
The number projected energy in this case is defined as
\begin{align}
&E(N) = \frac{1}{P(N)}{\text Re}\int_0^{\pi} \frac{d\eta}{\pi} e^{-i\eta N }{\cal E}[n(\xi,\xi |\eta),\ldots ],
\end{align}
in which one has to use the generalized densities in the expression for the energy density functional.
Mathematically this follows from the definition of a conditional probability and one has
\begin{align}
& E(N)=\frac{\int_0^{2\pi} d\eta \; e^{-iN\eta} \langle \Phi| \hat{H} |\Phi (\eta) \rangle 
                }{\int_0^{2\pi} d\eta \; e^{-iN\eta} \langle \Phi|\Phi (\eta) \rangle} = E_N, \\
&   \langle \Phi| \hat{H}|\Phi(\eta)\rangle = \sum_N E_N e^{i\eta N} |a_N |^2,  \nonumber\\
&   \langle \Phi |\Phi(\eta)\rangle = \sum_N e^{i\eta N} |a_N |^2,    |\Phi\rangle = \sum_N a_N|N\rangle, \nonumber    
\end{align}
if $[\hat{H},\hat{N}]=0$. In the sums degeneracies are implied and $|N\rangle$ 
are states with fixed particle number $N$ and average energy $E_N=\langle N|\hat{H}|N\rangle$. 
$E(N)$ can be evaluated with such a formula only if $a_N\neq0$,
thus iff the trial state $|\Phi\rangle$ has a non-vanishing overlap with the state $|N\rangle$. There is an 
immediate implication in these formulas
that within a DFT approach
${\cal E}[n(\xi,\xi |\eta),\ldots ]$ should be a faithful representation of $\langle \Phi| \hat{H} |\Phi (\eta) \rangle $.

In a similar manner one can evaluate any other number projected observables, or even 
combine particle and angular momentum projections for reaction fragments. In all the formulas
the $\Theta(z) {\textrm v}_n(\xi)$ components of the quasiparticle wavefunctions control the projected values of 
the observables, and thus, all matrix elements extend only over the matter distribution of the reaction fragments
in a well defined spatial region, once the reaction fragments are well separated. Since the overlap between
well separated fragments is vanishingly small, the formal non-commutativity between $\Theta(Z)$ and 
the angular momentum of a fragment $J_x$ with respect to its own center-of-mass is irrelevant.\\

{\bf Acknowledgements} \\

I thank L. Robledo, G. Bertsch, and M. Oi for input. This work was supported by U.S. Department of Energy,
Office of Science, Grant No. DE-FG02-97ER41014 and in part by NNSA
cooperative Agreement DE-NA0003841. 


\providecommand{\selectlanguage}[1]{}
\renewcommand{\selectlanguage}[1]{}

\bibliography{local_fission}

\end{document}